\documentclass[preprint,showkeys,preprintnumbers,amsmath,amssymb]{revtex4}

\usepackage{graphicx}
\usepackage{dcolumn}
\usepackage{bm}
\usepackage{epsfig} 
\usepackage{csquotes}

\begin{document}

\preprint{}

\title{Hopfield Network based Control and Diagnostics System for Accelerators}

\author{N. Joshi, O. Meusel, H. Podlech}
\email{njoshi@fias.uni-frankfurt.de}

\affiliation{%
Frankfurt Institute for Advanced Studies (FIAS),\\
600438 Frankfurt, Germany 
}
\affiliation[Also at]{Goethe University, Frankfurt am Main.}

\date{September 10, 2017}

\begin{abstract}

A recurrent artificial neural network known as Hopfield network is used for pattern storage.
Here we have applied this associative memory type network for pattern recognition for predictive controls and diagnostics in accelerator based systems. 
This system will be usefull for control systems and data acquisition system based on artificial intelligence at accelerator facilities.
In this publication we have discussed the role of Hopfield network as pattern recognition and auto encoder for denoising the images of ion beams obtained using CCD based optical systems.

\end{abstract}


\maketitle

\section{\label{sec:level1}Introduction\protect\\
 }

\begin{figure}[hhh]
    \centering
    \includegraphics[width=8cm, height = 6 cm]{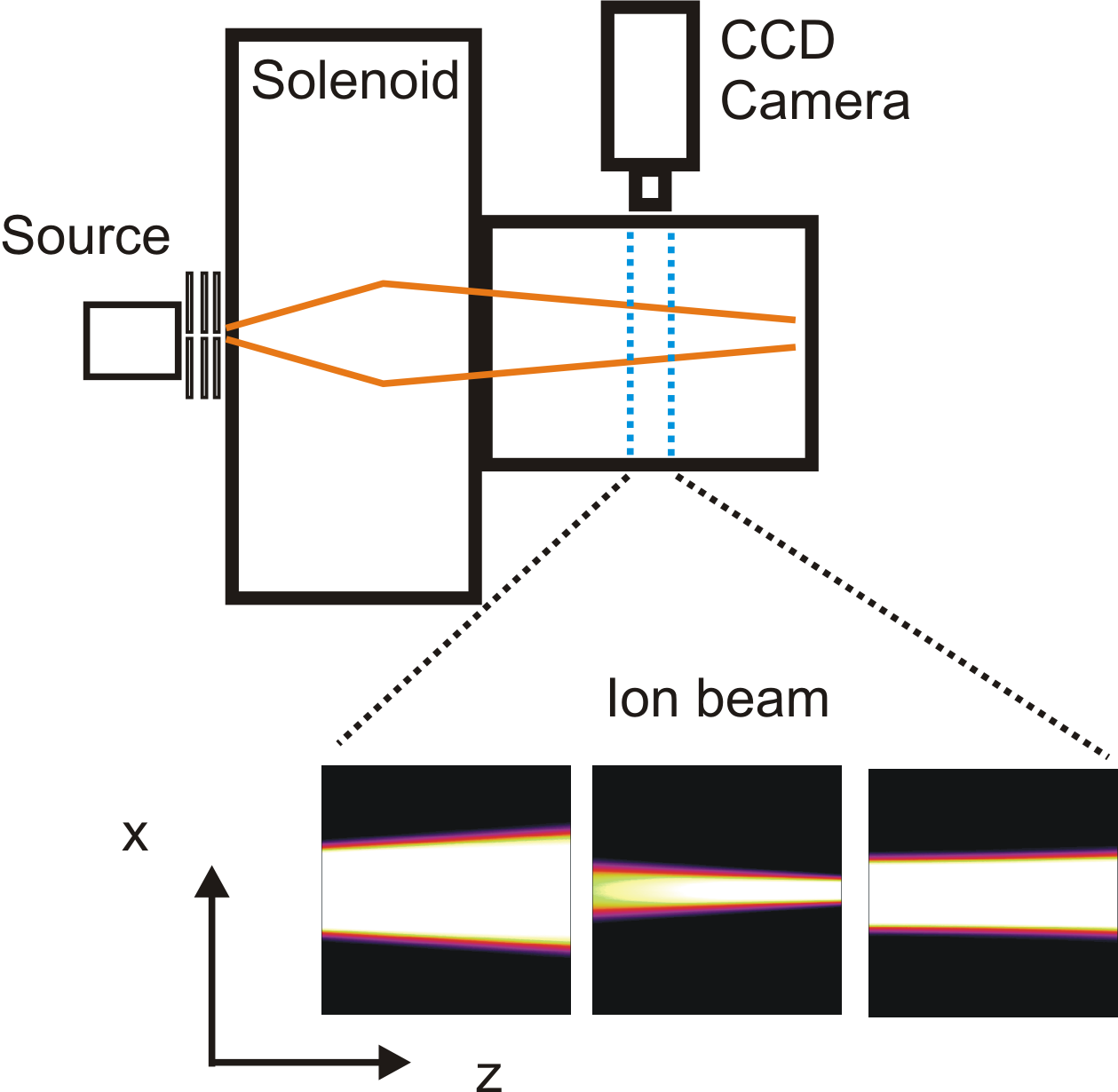}
    \caption{ Schematic for optical diagnostics.
      }
    \label{fig_1}
\end{figure}

With recent interests in machine learning and artificial intelligence in different fields, accelerator physicist also exploiting its applications \cite{Edelen}.
We have investigated the Hopfield network applied for controls and dignostics in the accelerator systems.
Here we consider a simple ion beam transport scheme; consisting of ion source, a focussing element (solenoid) and a diagnostic chamber with CCD images (see Fig \ref{fig_1}(top)).

\begin{figure}[hhh]
    \centering
    \includegraphics[width=2cm, height =2 cm]{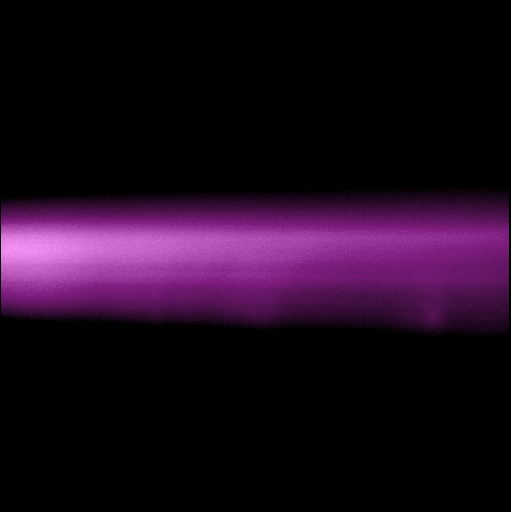}
    \includegraphics[width=2cm, height = 2 cm]{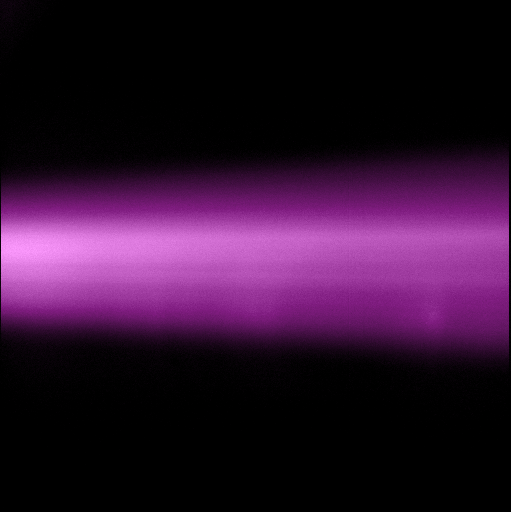}
    \includegraphics[width=2cm, height = 2 cm]{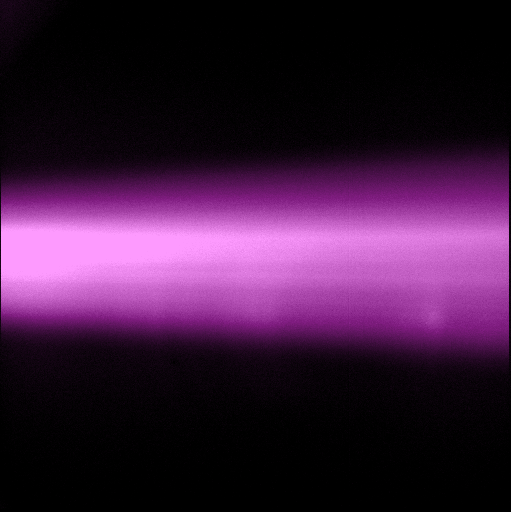} \\
\vspace{0.1cm}
    \includegraphics[width=2cm, height = 2 cm]{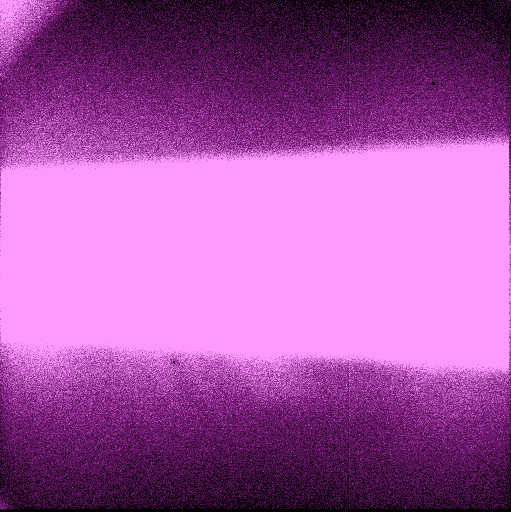}
    \includegraphics[width=2cm, height = 2 cm]{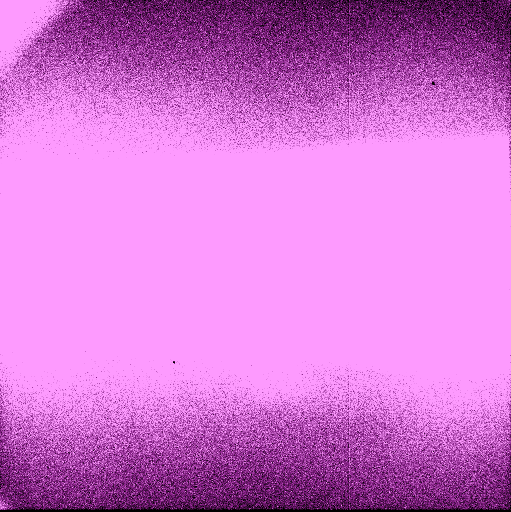}
    \caption{ Images representing beam loss due to vacuum breakdown (False colour representation)
      }
    \label{fig_2}
\end{figure}

The ion beam  transport  through solenoid is simulated considering constant initial distribution of ions for different magnetic field strengths. 
A CCD images are simulated considering its position tranverse to beam direction producing the images as shown in the Fig \ref{fig_1} (bottom).
Consider the scenario of beam loss due to vacuum breakdown. 
Fig \ref{fig_2} shows a few snapshots that represent this event.

\section{\label{sec:level1}Hopfield Network\protect\\
 }

Hopfield network is a type recurrent network  with binary threshold units i.e. units can take only two different values either $1$ or $-1$

\begin{figure}[hhh]
    \centering
    \includegraphics[width=6cm, height = 5 cm]{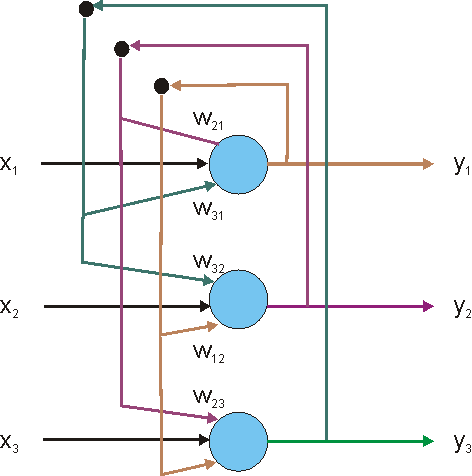}
    \caption{ Example of a Hopfield structure with three neuron.
      }
    \label{fig_3}
\end{figure}

Each pair of units $i$ and $j$ in the network have a connection and associated weights $w_{ij}$
Fig. \ref{fig_3} shows network with three neurons.
The connections in the network follow two conditions

\begin{itemize}
\item $w_{ii} $ = 0 : No self connection
\item $w_{ij} = w_{ji}$ : Symmetry
\end{itemize}

The updating of the node in the Hopfield network is given by

\begin{equation} \label{eq1}
\begin{split}
S_i & = +1 ~~~ if ~~  \sum_{j}{w_{ij}S_i} \ge \theta_{i}  \\
 & = -1  ~~~ otherwise
\end{split}
\end{equation}

where $w_{ij}$ connection strength between $i$ th and  $j$ th neuron, $S_i$ is the state of neuron $i$ and $\theta_i$ is the threshold of neuron $i$.
We have used asynchronous updating i.e. only one unit is updated randomly at a time.

\section{\label{sec:level1} Pattern recovery and predictive analysis in Hopfield Network   \protect\\
 }

\begin{figure}[hhh]
    \centering
    \includegraphics[width=6cm]{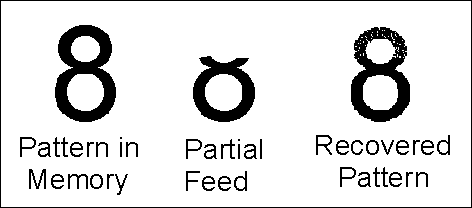}
    \caption{ Example of pattern recovery from Hopfield network. A figure of number eight is recovered from partial feed.
      }
    \label{fig_4}
\end{figure}

Fig. \ref{fig_4} demonstrates how Hopfield network is used for pattern recovery.
An image of figure $8$ is stored in the network or memory.
When partially completed image of the same is presented to the network its output is completed figure $8$.
Thus it is able to recall the correct pattern from the memory.

This type of network can be readily used as predictive system and efficient control system.

\begin{figure}[hhh]
    \centering
    \includegraphics[width=8cm]{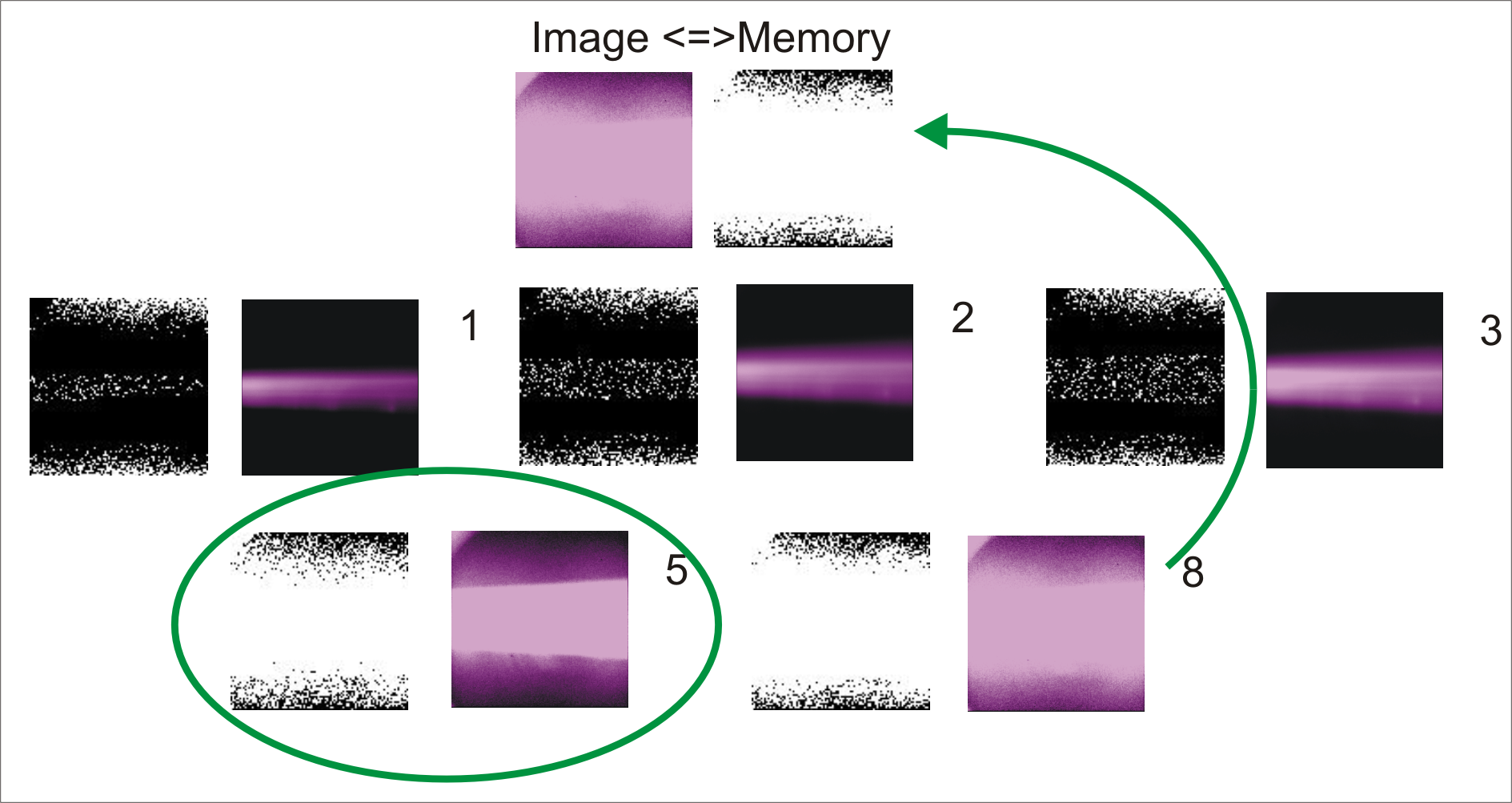}
    \caption{Sequence of images presented to the network. The last image was stored in memory for comparison. The network recognizes the pattern already at image number $5$ and predicts output completing image $8$.
      }
    \label{fig_5}
\end{figure}

The following example demostrates the Hopfield netwrok used as pattern necognition and prediction of vacuum breakdown.
Firstly, the time evolved images of beam loss due to vacuum breakdown were simulated. 
An image that can repesent vacuum breakdown; say image $8$ in this case, was presented to the network.
Typically $320 \times 320$ pixel image is converted in binary vector format and stored in the memory.
Please note here we refer to the memory of the network not the RAM or computer memory.
Then sequence of images were presented to the network.
Fig. \ref{fig_5} shows the output of the network as black and white image as a response from coloured image presented to it.

As one observe images $1, 2 ~and~ 3$ generate low respose but as the image $5$ is presented its response is the similar to that of the image $8$ which is in the memory. 
Thus the network has recalled the image from memory that corresponds to vacuum break down or that particular pattern.
For demonstration purpose only $8$ images are shown.

To compare two pictures,  let us calculate similarity between two images.
The grey-scale image is normalized and an accurate $2D$ correlation is performed, i.e the correlation on two input matrices.
The 2D Correlation performs 2D correlation on two input matrices.
Both linear and circular correlation can be computed.
Two computation methods are available: a fast algorithm based on FFT and an accurate method based on shift accumulation.
With a normalized result, it is easier to tell the degree  to which the two input signals are correlated.

 \begin{figure}[hhh]
    \centering
    \includegraphics[width=6cm]{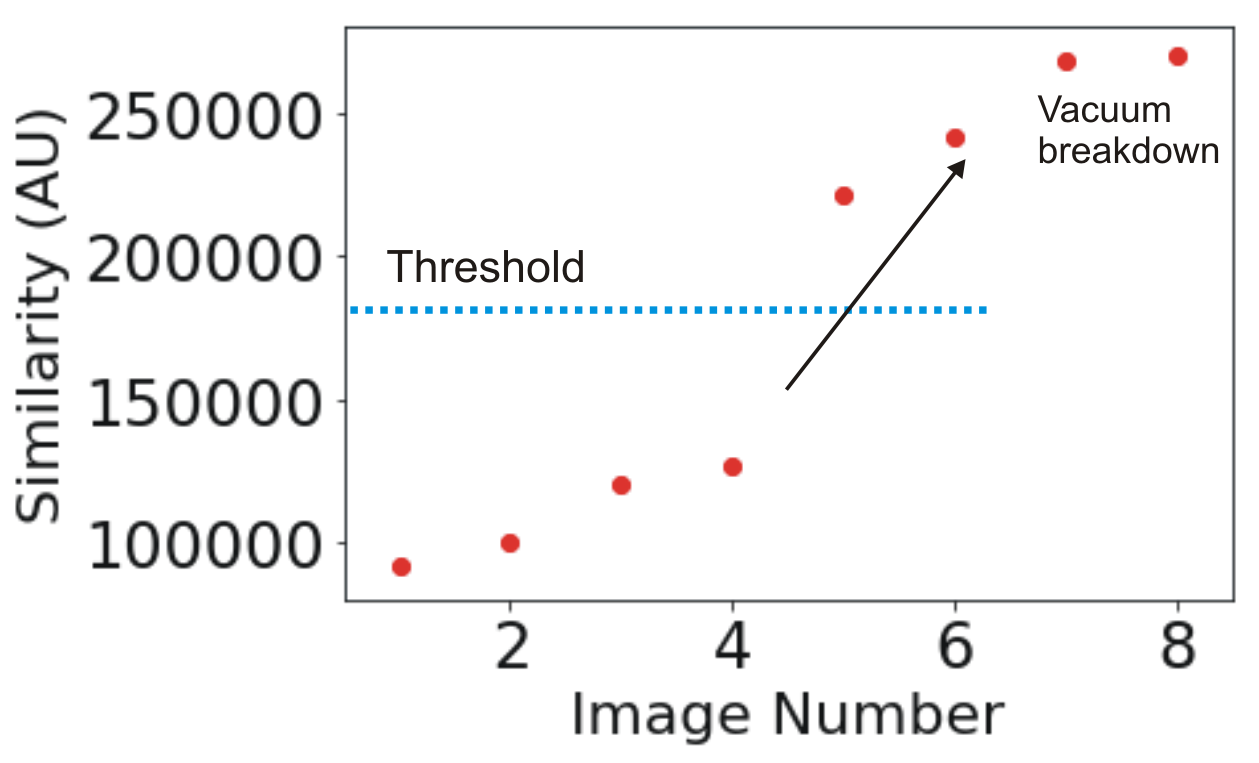}
    \caption{Evolution of similarity between pictures.
      }
    \label{fig_6}
\end{figure}

Fig. \ref{fig_4} shows the Similarity number as a function of image number. 
Images $1$ to $4$ remains low but after image $5$ the similarity jumps up and moves to 
By defining a suitable threshold value one can build control feedback.
Thus Hopfield network based algorithm along with optical diagnostics can be integrated in control system.

As we think more in details about this method, only similarity between patterns will not suffice as a foolproof solution.
Thus instead of similarity measure, the cascade network is used which is fed with the pattern that represents variation of similarity over the time and anomaly is detected as an abrupt jump in the time evolution of the images.

\section{\label{sec:level1}Autoencoder and Denoising  \protect\\
 }

As its known the real world images from experiments wont be so clear.
Some sort of noise will always be present in data acquisition system.
To overcome this issue the neural network is used in auto encoder mode.
In this format multiple trainning images from ion beam are used. 
The trainning dataset contains images with distortion by adding artifitial noise.
After training, the network is presented with new image it is able recognize the signal i.e. ion beam from the backgound noise.

 \begin{figure}[hhh]
    \centering
    \includegraphics[width=6cm]{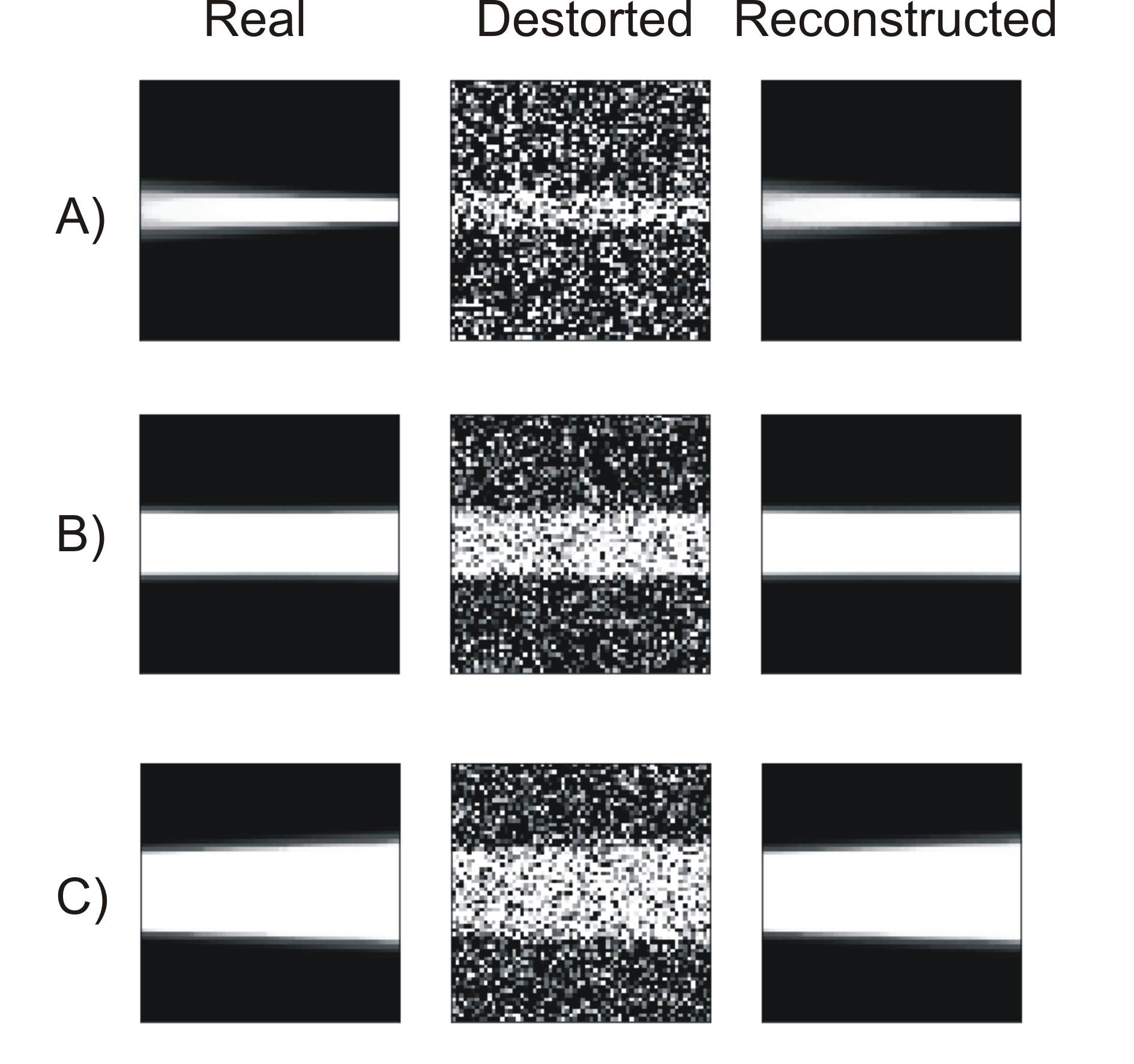}
    \caption{ Example of noise cancellation using neural network. The images are added with noise which are then reconstructed back using neural network.
      }
    \label{fig_6}
\end{figure}

Fig. \ref{fig_6} shows the example images and its distroted version with noise addition. 
Three examples shows a focussed ion beam, parallel beam and a slightly diverging beam. 
The third column shows the reconstructed image.
The efficiency of the reconstructed image depends on the noise level.

\begin{figure}[hhh]
    \centering
    \includegraphics[width=6cm]{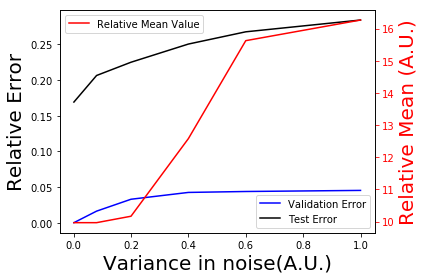}
    \caption{Errors and Entropy Gain as a function of increasing noise. In this particular example Gaussian noise is added with different variance or the sigma value.
      }
    \label{fig_8}
\end{figure}

Fig. \ref{fig_8} shows the efficiency dependent on the noise level in terms of  Relative error and relative mean of the images.

It shows that as the noise level is increased the image can not be differentiated and the relative mean value increases.
In this example we have consider only gaussian type of noise.
%
%
%

\section {Conclusions}
\label {Conclusions}

Here we have presented Hopfield network based algorithm tha can be used for control systems.
In complete scenario autoencoders will be used to clear up the noisy images. 
The online system is then run with pattern recognition which in turn can be used either for predictive analysis or just as data acquisition.
These codes were implemented in Python along with standard packages  like Matplotlib, scikit-image etc., freely available under open source licence.

\end{document}